\documentclass{aa} 
\usepackage{times}
\usepackage{color}
\usepackage{comment}
\usepackage{natbib}
\usepackage{url}
\usepackage{hyperref}
\usepackage{subfigure}
\usepackage{xspace}
\usepackage[percent]{overpic}
\usepackage{subfloat}
\usepackage{subcaption}
\usepackage{graphicx}
\usepackage{caption}
\usepackage{subcaption}
\usepackage{afterpage}
\usepackage{float}
\usepackage{multicol} 
\usepackage{adjustbox} 
\usepackage{longtable}
\usepackage{gensymb}
\usepackage{afterpage}
\usepackage{pdflscape}
\usepackage{txfonts}

\newcommand{\nustar}{{\it NuSTAR}\xspace}
\newcommand{\bat}{{\it Swift}/BAT\xspace}
\newcommand{\hxmt}{{\it Insight}-HXMT\xspace}

\newcommand{\src}{Vela~X-1\xspace}
\graphicspath{{./}{figures/}}

\begin{document}

    \title{Long-term evolution of cyclotron resonant scattering features in the accreting pulsar \src: A pulse-to-pulse approach}

    \author{Yu-Jia~Du \inst{\ref{in:Tub}}
          \and Peng-Ju~Wang \inst{\ref{in:Tub}}
          \and Lorenzo~Ducci \inst{\ref{in:Tub},\ref{in:Genf}}
          \and Long~Ji \inst{\ref{in:SYSU}}
          \and Ling-Da~Kong \inst{\ref{in:Tub}}
          \and Qing-Cui~Bu \inst{\ref{in:CCNU}}
          \and Chirag~Mehrotra \inst{\ref{in:Tub}} 
          \and Andrea~Santangelo \inst{\ref{in:Tub}}
          }

   \institute{Institut f\"ur Astronomie und Astrophysik, Universit\"at T\"ubingen, Sand 1, D-72076 T\"ubingen, Germany \label{in:Tub}\\
   \email{du@astro.uni-tuebingen.de}
   \and
   ISDC Data Center for Astrophysics, Université de Genève, 16 chemin d’Écogia, 1290, Versoix, Switzerland \label{in:Genf}
   \and
       School of Physics and Astronomy, Sun Yat-Sen University, Zhuhai, 519082, People's Republic of China \label{in:SYSU}
    \and
       {Institute of Astrophysics, Central China Normal University, Wuhan 430079, P.R. China \label{in:CCNU}}
}

\abstract
{
{We investigated the long-term evolution of the cyclotron line energy, as well as the relationship between cyclotron line energy and luminosity in the high-mass X-ray binary \src, based on archival \bat monitoring from 2005 to 2024 and pulse-to-pulse analysis of nine \nustar observations from 2012 to 2024. Our results provide the first confirmation that the long-term decay of the harmonic line energy ($E_{\rm cyc,H}$) in \src has ended. We further report the first detection of a transient increase in $E_{\rm cyc,H}$ between 2020 and 2023, which suggests a sudden and significant change in the magnetic field configuration or accretion geometry. In addition, $E_{\rm cyc,H}$ shows slightly lower values at low luminosities and tends to flatten at higher luminosities, in the range of $(0.13\text{--}1.21) \times {10}^{37}$~erg~s$^{-1}$. The fundamental line energy ($E_{\rm cyc,F}$) exhibits no significant variation with time or luminosity, remaining stable at approximately 25~keV.}
}

\keywords{X-rays:binaries --
            pulsars: individual: Vela X-1
            }

\maketitle
%

\section{Introduction}\label{sec:introduction}

Vela X-1 is an eclipsing high-mass X-ray binary (HMXB) and one of the earliest X-ray sources discovered in the history of X-ray astronomy \citep{1967ApJ...150...57C}. It consists of the B0.5~Ib supergiant HD~77581 \citep{1972ApJ...175L..19H}, and a wind-accreting neutron star (NS) with a spin period of $\sim$283~s, rotating in an 8.9~d orbit around the companion.  The distance to the system is $1.9 \pm 0.2$~kpc \citep{1985ApJ...288..284S}. The mean luminosity is 5 $\times$ $10^{36}\,$erg\,$\rm{s}^{-1}$ \citep{2010A&A...519A..37F}. A detailed description of the system can be found in the recent review by \citet{2021A&A...652A..95K}.

Some highly magnetized accreting NSs show absorption-like lines in their high-energy X-ray spectra, called cyclotron lines or cyclotron resonant scattering features (CRSFs). They are produced in strong magnetic fields, where electrons are quantized onto Landau levels. Photons with energies close to the Landau levels are removed from the observed X-ray spectrum by scattering off these electrons. CRSFs provide a direct measurement of the magnetic field. The centroid energy of the CRSFs can be expressed as 
\begin{equation}
\label{equ:Bcyc}
E_{\rm cyc} \approx \frac{n}{(1+z)} 11.6~[{\rm keV}]  \times B_{12},
\end{equation}
where $n$ is the number of Landau levels, $B_{12}$ is the magnetic field strength in units of \(10^{12}\) gauss, and $z$ is the gravitational redshift of the line formation region \citep{2019A&A...622A..61S}.

Based on observations with the High Energy X-ray Experiment (HEXE), \citet{1992fxra.conf...51K} reported the first evidence of CRSFs in \src: a fundamental line at approximately 25~keV and a first harmonic line near 50~keV. Subsequent studies confirmed the presence of both features, with the fundamental line typically observed at $\sim$23--27~keV and the harmonic at $\sim$45--54~keV \citep{1997A&A...325..623K, 2002A&A...395..129K, 2007ESASP.622..479S, 2013ApJ...763...79M, 2013ApJ...767...70O}. Notably, the fundamental line is considerably weaker than the harmonic and is not always detectable. 
Significant variability has been observed with the CRSF in \src, regarding its centroid energy $E_{\rm cyc}$ and other characteristic parameters (e.g., width and optical depth). These parameters vary with luminosity and with time.
\citet{2014AAS...22343820F} detected both CRSFs based on \nustar observations, and for the first time reported a positive correlation between the energy of the harmonic line ($E_{\rm cyc,H}$) and luminosity, consistent with theoretical predictions in the subcritical accretion regime \citep{2012A&A...544A.123B}. In contrast, the energy of the fundamental line ($E_{\rm cyc,F}$) exhibits a more complex, nonmonotonic dependence on luminosity. This positive $E_{\rm cyc,H}$–luminosity correlation was further confirmed by \citet{2016MNRAS.463..185L} through an analysis of long-term \bat data. In addition, they discovered a secular decline in $E_{\rm cyc,H}$ at a rate of $\sim$0.36~keV~yr$^{-1}$ between 2004 and 2010. This establishes \src as the second known source, after Her X-1 (see the summary by \citealt{2020A&A...642A.196S}), to exhibit a long-term $E_{\rm cyc}$ decay. {\citet{2019MNRAS.484.3797J} confirmed that the decay persisted until 2012 and reported that $E_{\rm cyc,H}$ remained stable thereafter based on continued \bat monitoring.}

Two key questions are whether the long-term decay of $E_{\rm cyc,H}$ in \src has truly ended and whether there is any indication of a subsequent increase. However, since the observations were obtained at different luminosity levels, the $E_{\rm cyc}$–luminosity correlation presents a major challenge for disentangling intrinsic long-term evolution of $E_{\rm cyc}$ from its luminosity dependence. Previous studies have addressed this issue by using normalization techniques in Her X-1 \citep{2014A&A...572A.119S}, and by comparing evolutionary trends across different epochs in \src \citep{2016MNRAS.463..185L}. For this work our aim was to address this issue by applying a pulse-to-pulse analysis technique.

{For this paper we analyzed \bat monitoring data of \src spanning from January 2005 to December 2024; we employed the methodology and software developed by \citet{2015A&A...578A..88K} and validated by \citet{2019MNRAS.484.3797J}. Here we present a detailed spectral analysis using nine \nustar observations conducted between 2012 and 2024.} The structure of the paper is as follows. In Section~\ref{sec:data_analysis}, we describe the observations and data reduction procedures. Section~\ref{sec:results} presents the results of our analysis. In Section \ref{sec:discussion} and \ref{sec:conclusions}, we discuss and summarize our results. Additional results are included in the Appendix.

\begin{table*}[!t]
\caption{Details on \nustar observations of \src\ between 2012 and 2024. \label{tab:ObsID}} 
\centering
\begin{tabular}{lccrc}
\hline\hline 
ObsID&Obs. Time&Duration&Exposure&Period\\
     &YYYY-MM-DD&MJD&\multicolumn{1}{c}{s}&s \\
\hline	
10002007001 & 2012-07-09  & 56117.63-56117.94 & 10813 & 283.48 \\
30002007002 & 2013-04-22  & 56404.53-56404.84 & 7137 & 283.41 \\
30002007003 & 2013-04-22  & 56404.86-56405.78 & 24496 & 283.42 \\
90402339002 & 2019-01-10  & 58493.18-58494.09 & 36036 & 283.45 \\
30501003002 & 2019-05-03  & 58606.87-58608.25 & 40375 & 283.43 \\
90602328002 & 2020-09-24  & 59116.98-59117.22 & 10728 & 283.53 \\
90602328004 & 2020-09-26  & 59119.00-59119.22 & 9593 & 283.43 \\
90602328006 & 2020-09-29  & 59121.68-59121.91 & 8977 & 283.46 \\
91002349002 & 2024-11-19  & 60633.05-60633.96 & 36930 & 283.57 \\

\hline
\end{tabular}
\end{table*}

\section{Observations and data analysis}\label{sec:data_analysis}

\subsection{\bat}

In this work, we include publicly available archival data obtained with the Burst Alert Telescope (BAT; \citealt{2005SSRv..120..143B}), a hard X-ray detector on board the Neil Gehrels {\it Swift} observatory \citep{2004ApJ...611.1005G}. The aim of the BAT spectral analysis is to investigate the evolution of the energy of the harmonic cyclotron line. Due to the regular monitoring of the source by BAT, the dataset is nearly uniformly sampled with minimal time gaps. The instrument is sensitive in the energy range of 15--150~keV and is suited for the study of the $\sim$55~keV harmonic cyclotron line in \src.

The analysis covers data collected from January 2005 through December 2024, all of which are accessible via HEASARC.\footnote{\url{https://heasarc.gsfc.nasa.gov/docs/archive.html}} The selected data were obtained in survey mode, in which events were collected in the detector plane histograms (DPHs) accumulated over a five-minute exposure time. 
The data reduction follows the method described by \citet{2015A&A...578A..88K} and \citet{2019MNRAS.484.3797J}. 
The spectra were generated using the \texttt{batbinevt} tool from the \texttt{HEASoft} v6.34. We extracted the spectra only if the source could be identified in the sky map. The recommended BAT systematic uncertainties were applied using the \texttt{batphasyserr} task. Corrections to the BAT energy scale for detector nonlinearities and detector-dependent offsets not accounted for by onboard calibration were performed using the \texttt{baterebin} tool.

To enhance the statistics in the spectral analysis, we jointly fitted hundreds of spectra within three-month intervals; the spectral shapes were assumed to be the same, while the normalizations were variable. The validity of this method was previously verified by \citet{2015A&A...578A..88K} and \citet{2019MNRAS.484.3797J}.

\subsection{\nustar}

The Nuclear Spectroscopic Telescope Array (\nustar) is the first focusing hard X-ray observatory and is equipped with two identical co-aligned X-ray telescopes, Focal Plane Module A and B (FPMA and FPMB; \citealp{2013ApJ...770..103H}). The instruments cover a wide energy range of 3--79~keV and provide an imaging resolution of 18\arcsec at full width at half maximum (FWHM) and a spectral energy resolution of 400~eV (FWHM) at 10~keV. 
In Table \ref{tab:ObsID} we list nine observations of \src performed by \nustar over the time period 2012 to 2024. We used the NUSTARDAS pipeline v2.1.2 and \texttt{HEASoft} v6.33.2 with \nustar CALDB v20240520. As suggested by \citet{2022A&A...660A..19D}, we used the old FPMA ARF\footnote{\url{https://nustarsoc.caltech.edu/NuSTAR_Public/NuSTAROperationSite/mli.php}} for observation 30501003002 to avoid differences between the two focal plane modules, due to an overly low-energy effective area correction.
The event times are corrected for barycentric motion using the \texttt{barycorr} tool from NUSTARDAS and the binary orbit using the ephemeris from the Fermi GBM website.\footnote{\url{https://gammaray.nsstc.nasa.gov/gbm/science/pulsars/lightcurves/velax1.html}} For each observation, we extracted source spectra from a circular region with a 90 arcsec radius and background spectra from a circular region with a 60 arcsec radius. We performed epoch folding \citep{1987A&A...180..275L} on the extracted light curves to measure the spin period; the results are presented in Table \ref{tab:ObsID}. 

We wanted to investigate the energy dependence on luminosity across different time intervals to explore the long-term evolution of spectral variability. However, the available data were insufficient to establish a definitive correlation. To address this limitation, we subsequently performed a pulse-to-pulse analysis of the \nustar\ observations.
This technique, originally introduced by \citet{2011A&A...532A.126K}, effectively expands the observed luminosity range and provides more detailed insights into spectral evolution. The analysis is described in detail in Section~\ref{sec_p2p}.

Both average and pulse-to-pulse spectra were grouped to ensure a minimum of 30 photons per energy bin. Spectral fitting was performed over the 3--70~keV energy range. All quoted uncertainties on spectral parameters correspond to the 68\% confidence level.

\section{Results}\label{sec:results}

\subsection{\bat result}

\begin{figure}[H]
\centering
\includegraphics[width=9cm]{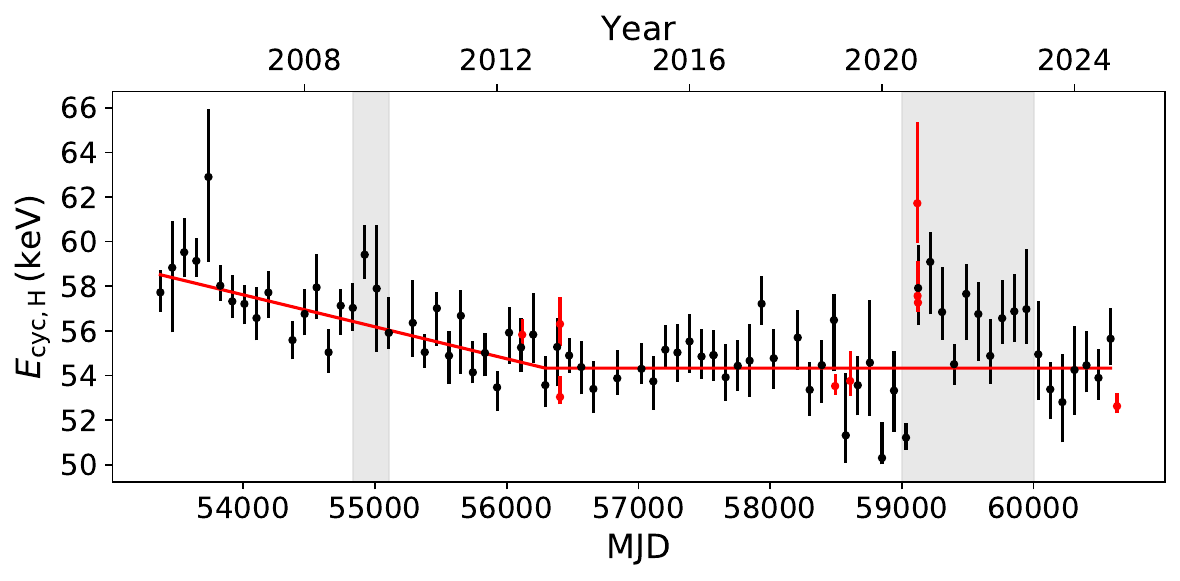}
\caption{Long-term evolution of the centroid energy of the harmonic cyclotron line in \src as observed with \bat. The gray shaded regions indicate epochs of apparent $E_{\rm cyc,H}$ humps and the red points represent the time-averaged results from \nustar observations in Section \ref{nu_average}.}
\label{BAT_Eh}
\end{figure}

This study includes \bat data spanning from January 2005 to December 2024. The data analysis procedures closely follow those described by \citet{2015A&A...578A..88K}, \citet{2016MNRAS.463..185L}, and \citet{2019MNRAS.484.3797J}.
{To model the continuum of Vela X-1, we tested different spectral shapes. A simple power law modified with an exponential cutoff (\texttt{cutoffpl}) does not fit the data. A \texttt{highecut} model, another often-used model for the continuum in accreting neutron stars, is also not adequate to describe the data. We adopted the \texttt{comptt} model to describe the continuum, consistent with previous analyses of BAT data \citep{2016MNRAS.463..185L, 2019MNRAS.484.3797J}. The fundamental line at approximately 25~keV is not detectable by BAT. 
} 
The combined model, \texttt{comptt $\times$ gabs}, yields a good fit across all spectra. The temperature of the seed photons was fixed at 1~keV during the fitting. The resulting parameters are available in electronic form at the CDS, and the corresponding results are illustrated in Fig.~\ref{BAT_Eh}.
{In Figure \ref{BAT_Eh}, the black data represents the centroid energy of the harmonic cyclotron line as derived by the BAT data, while the red data represent the time-averaged results from the \nustar observations discussed in Section \ref{nu_average}.}
Initially, the line energy exhibited a significant decrease and remained nearly constant thereafter. A pronounced fluctuation occurred between MJD 59000 and MJD 60000 (corresponding to the years 2020 to 2023).
After this period, the energy returned to its pre-fluctuation level.
We adopted a broken linear function to model the evolution of $E_{\rm cyc,H}$, following the approach of \citet{2019MNRAS.484.3797J}:
\begin{equation}
E_{\rm cyc,H}(t) =
\begin{cases}
E_0 + a \times (t - t_0) & t \leq t_{\mathrm{crit}}
\\
\text{const} = E_0 + a \times (t_{\mathrm{crit}} - t_0)& t > t_{\mathrm{crit}}
\end{cases}
\end{equation}
Here $E_0$ is the reference energy at fixed epoch $t_0$=53371 (MJD), $a$ is the linear slope describing the decreasing trend, {and $t_{\mathrm{crit}}$ is the break time after which $E_{\mathrm{cyc,H}}$ remains constant.}
The resulting $t_{\mathrm{crit}}$ is MJD $56293\pm264$. The decrease rate of $E_{\rm cyc,H}$ is $-0.584 \pm 0.073~\mathrm{keV~yr^{-1}}$. After the break, the $E_{\rm cyc,H}$ remains constant at $54.60\pm0.83 ~\mathrm{keV}$. Over the evolution, two abrupt humps are apparent, occurring around MJD 55000, and between MJD 59000 and MJD 60000.

\subsection{\nustar result}
\subsubsection{The average spectra result}
\label{nu_average}

\begin{figure}
\centering
\includegraphics[width=9cm]{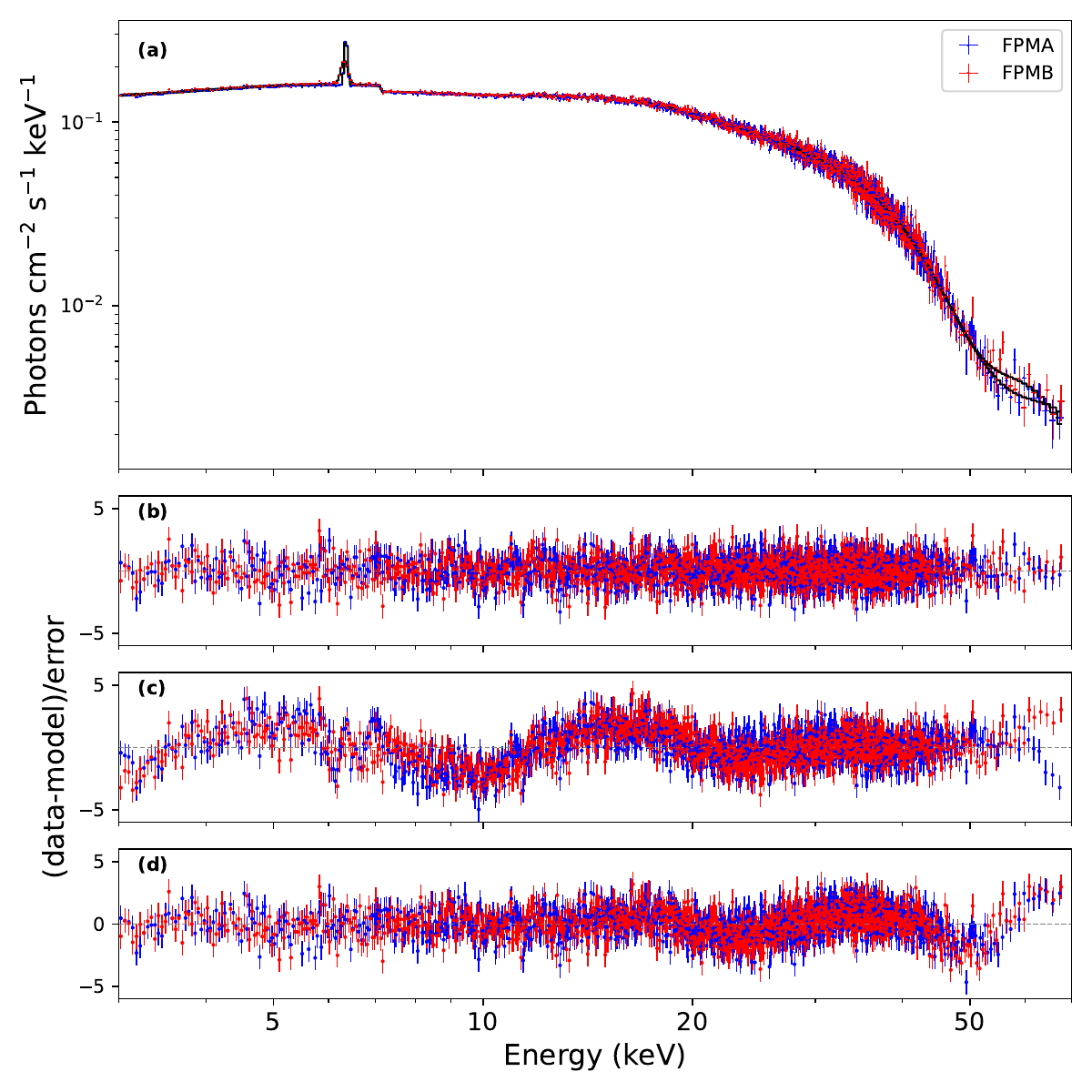}

\caption{(a) Spectrum and best-fit model for ObsID 91002349002 in energy range 3--70~keV using the model \texttt{const$\times$tbabs$\times$pcfabs(gauss+gauss+gabs$\times$gabs$\times$fdcut)}. The FPMA data is in red and the FPMB data in blue. The best-fit model is shown in black. (b) Residuals for the best-fit model. (c) Residuals after fitting without the feature around 10~keV. (d) Residuals after fitting without the CRSFs.}
\label{average_spectrum}
\end{figure}

\begin{table} 
\caption{Best-fit parameters of the phase-averaged spectrum for ObsID 91002349002 using the model \texttt{const$\times$tbabs$\times$pcfabs(gauss+gauss+gabs$\times$gabs$\times$fdcut)} in the 3--70~keV energy band. 
} \label{table:averaged_par}
\centering
\vspace{-0.2cm}
\renewcommand{\arraystretch}{1.35}
\begin{tabular}{lc}
\hline\hline 
Parameter &  Value\\ 
 \hline 
$C_{\rm FPMA}$ (fixed) & 1 \\
$C_{\rm FPMB}$ & $1.0100_{-0.0005}^{+0.0013}$ \\
$N_{\textrm{H1}}$ [$10^{22}\,$cm$^{-2}$] (fixed)& 0.371 \\
$N_{\textrm{H2}}$ [$10^{22}\,$cm$^{-2}$]& $20.8_{-0.9}^{+1.1}$ \\
CF & $0.30_{-0.01}^{+0.01}$ \\
$E_{\rm 10~keV}\,$[keV] & $8.6_{-0.2}^{+0.2}$ \\
$\sigma_{\rm 10~keV}\,$[keV] & $3.1_{-0.2}^{+0.2}$ \\
Norm$_{\rm 10~keV}$ & $-0.014_{-0.003}^{+0.002}$ \\
$E_{\rm K\alpha}\,$[keV] & $6.341_{-0.005}^{+0.005}$ \\
$\sigma_{\rm K\alpha}\,$[keV] & $0.051_{-0.036}^{+0.014}$ \\
Norm$_{\rm K\alpha}$ & $0.00175_{-0.00008}^{+0.00005}$ \\
$E_{\rm cyc,F}\,$[keV] & $24.55_{-0.48}^{+0.48}$ \\
$\sigma_{\rm cyc,F}\,$[keV] & $0.5\times\sigma_{\rm cyc,H}$ \\
$d_{\rm cyc,F}$ & $0.46_{-0.07}^{+0.14}$ \\
$E_{\rm cyc,H}\,$[keV] & $52.63_{-0.32}^{+0.57}$ \\
$\sigma_{\rm cyc,H}\,$[keV] & $7.24_{-0.38}^{+0.64}$ \\
$d_{\rm cyc,H}$ & $14.98_{-1.44}^{+2.52}$ \\
$E_{\rm cut}\,$[keV] & $22.2_{-1.3}^{+0.9}$\\
$E_{\rm fold}\,$[keV] & $10.8_{-0.2}^{+0.4}$ \\
$\Gamma$ & $1.08_{-0.02}^{+0.01}$ \\
Norm$_\Gamma$ & $0.267_{-0.005}^{+0.008}$ \\
Flux$_{\rm 3-79~keV}$ & $6.592_{-0.014}^{+0.014}$ \\
$\chi^2$/d.o.f. & $555.27/448$ \\

\hline
\end{tabular}
\vspace{-0.2cm}
\tablefoot{All reported errors are at the 68\%  confidence level and based on the MCMC chain values.
Normalization of the power law in units of photon\,cm$^{-2}$\,s$^{-1}$~keV$^{-1}$ at 1~keV.
Unabsorbed flux (in units of $10^{-9}\,$erg\,cm$^{-2}$\,s$^{-1}$) calculated for the entire model, obtained using the \texttt{cflux} command from \textsc{xspec}. Uncertainties are given for a 68\% confidence level.
}
\end{table}

{For the \nustar observations, the Fermi–Dirac cutoff model (FDcut; \citealp{1986LNP...255..198T}) provides a good fit to the continuum and was also adopted in previous studies \citep{2014AAS...22343820F,2016MNRAS.463..185L,2022A&A...660A..19D}.} The FDcut model, characterized by the photon index $\Gamma$, cutoff energy $E_{\mathrm{cut}}$, and folding energy $E_{\mathrm{fold}}$, takes the form:
\begin{equation}
F(E) \propto E^{-\Gamma} \left(1 + \exp\left( \frac{E - E_{\mathrm{cut}}}{E_{\mathrm{fold}}} \right)\right)^{-1}.
\end{equation}
The two CRSFs are well described with two Gaussian absorption components (\texttt{gabs}). In all the spectra, we detect with high significance the fundamental at $\sim25$~keV and the second harmonic at $\sim$ 52--55~keV. We constrained the width of the fundamental line to be half that of the harmonic line, $\sigma_{\mathrm{cyc,F}} = 0.5 \times \sigma_{\mathrm{cyc,H}}$, following the approach adopted by \citet{2014AAS...22343820F} and \citet{2022A&A...660A..19D}. This assumption is motivated by the physical interpretation that harmonic lines originate from resonant scattering between the ground Landau level and higher excited levels (see equation \ref{equ:Bcyc}). Consequently, the energies of the harmonic lines are expected to be twice that of the fundamental line, and their width scale accordingly. We adopted the \texttt{tbabs} absorption model, and set the abundances to \texttt{wilm} \citep{2000ApJ...542..914W}. A single-absorption model cannot accurately describe the spectrum at lower energies because of the contribution of the absorption from the stellar wind. Therefore, we added a partial covering model \texttt{pcfabs} to take into account the wind structure \citep{2014AAS...22343820F,2022A&A...660A..19D}.The interstellar absorption component ($N_{\mathrm{H1}}$) was fixed at $0.371 \times 10^{22}\,\mathrm{cm}^{-2}$, as determined from NASA’s HEASARC $N_{\mathrm{H}}$ tool.\footnote{\url{https://heasarc.gsfc.nasa.gov/cgi-bin/Tools/w3nh/w3nh.pl}} As \nustar does not provide coverage below 3~keV, it is not well suited to constraining low absorption values. In contrast, the absorption due to the stellar wind ($N_{\mathrm{H2}}$) was left as a free parameter. A Gaussian emission line centered around $\sim6.4$~keV was added to account for iron fluorescence.  In addition, a broad Gaussian component around $\sim10$~keV was included, as this feature is essential for an adequate spectral fit. Similar features have been reported in previous studies of \src \citep{2014AAS...22343820F, 2022A&A...660A..19D}, and in other sources, where they typically appear near the spectral cutoff energy \citep{2012A&A...547A.103N, 2013ApJ...762...61D, 2021A&A...652A..89N}. In the case of Vela X-1, this feature has not been interpreted as a CRSF, although its physical origin remains uncertain. To enable joint fitting of the two instruments, a cross-normalization constant was applied. Throughout the analysis, the \texttt{cflux} convolution model was used to derive all unabsorbed flux values.

\begin{figure}
\centering
\includegraphics[width=9cm]{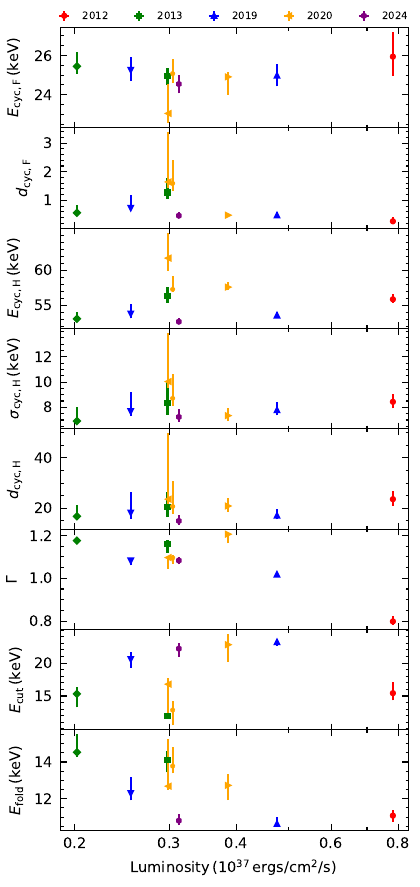}

\caption{Flux-resolved spectral parameters derived from nine \nustar observations in the 3--70~keV energy range, shown as a function of intrinsic source luminosity. From top to bottom: Energy of the fundamental line, energy of the harmonic line, width of the harmonic line, strength of the harmonic line, photon index, cutoff energy, and folding energy.}

\label{average_par}
\end{figure}

We present the results of ObsID 91002349002 as a representative example; this dataset has not been analyzed in previous studies. {Figure \ref{average_spectrum} shows the broadband spectrum. Panel (c) presents the residuals when the $\sim$10 keV component is excluded from the model, and Panel (d) presents the residuals when the two CRSFs are excluded.} Table~\ref{table:averaged_par} summarizes the spectral parameters obtained from the best-fit model for this observation. The model yielded a $\chi^2$ value of 555.27 for 448 degrees of freedom. The best-fit cyclotron line energies were $E_{\mathrm{cyc,F}} = 24.55_{-0.48}^{+0.48}$~keV and $E_{\mathrm{cyc,H}} = 52.63_{-0.32}^{+0.57}$~keV. This model was subsequently applied to fit the average spectra of all \nustar observations. The resulting parameters are available in electronic form at the CDS. Figure~\ref{average_par} shows the best-fit spectral parameters, including the energy and strength of the fundamental line; the energy, width, and strength of the harmonic line; and the continuum parameters, as functions of the intrinsic source luminosity (3--70~keV) assuming a distance of 1.9~kpc. Since the measured cyclotron line energies can be affected by the continuum shape and the line widths, we also constructed corner plot of the distribution for all the parameters (see Figure \ref{Corner_plot} in the Appendix).

The luminosity range derived from the time-averaged spectral analysis is approximately $ (0.2\text{--}0.8) \times 10^{37}$\,erg\,s$^{-1}$. We observed a sudden drop in $E_{\rm cut}$ at around $0.3 \times 10^{37}$\,erg\,s$^{-1}$. For the three corresponding observations, we fixed $E_{\rm cut}$ at 20~keV. This adjustment has no significant impact on the parameters of either the harmonic or the fundamental cyclotron lines. The observed drop is likely due to relatively low statistics in these spectra. No clear trend is observed in the evolution of $E_{\mathrm{cyc,F}}$ with luminosity; the values range from a minimum of $23.41_{-0.14}^{+0.86}$~keV to a maximum of $25.94_{-0.97}^{+1.25}$~keV. Similarly, no obvious correlation between $E_{\mathrm{cyc,H}}$ and luminosity can be established based on the nine averaged measurements. {The three 2020 observations show relatively high values of $E_{\mathrm{cyc,H}}$, reaching $61.72_{-1.78}^{+3.64}$~keV (ObsID 90602328002), $57.57_{-0.45}^{+0.73}$~keV (ObsID 90602328004), and $57.27_{-0.40}^{+1.86}$~keV (ObsID 90602328006), consistent with the increase in the line energy observed with the \bat\ monitoring since 2020.}
To more accurately trace the evolution of the line energy, we employed a pulse-to-pulse analysis technique.

\subsubsection{The pulse-to-pulse analysis}\label{sec_p2p}

\begin{figure}
\centering
\includegraphics[width=9cm]{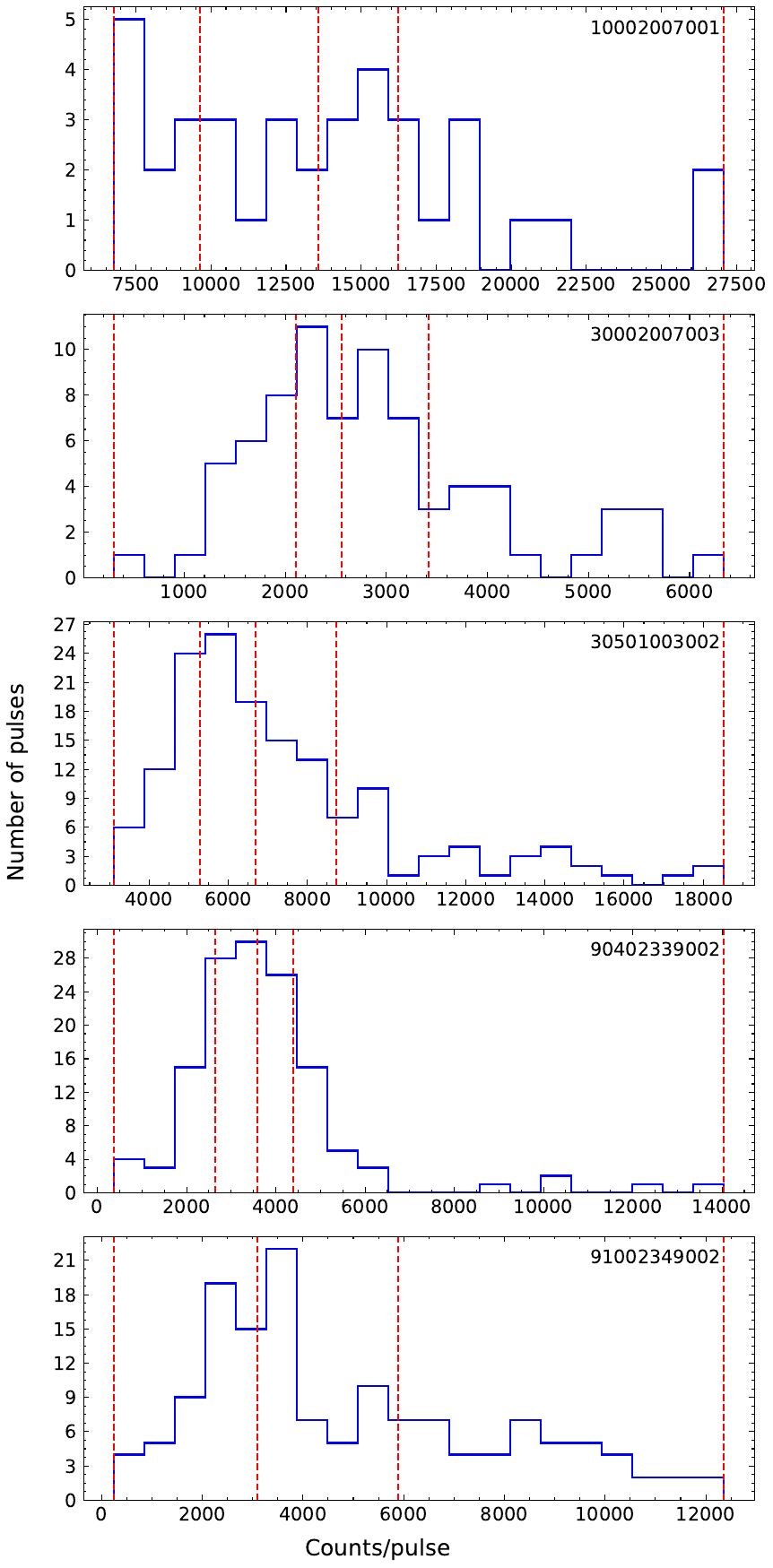}
\caption{Distributions of counts per individual pulse (i.e. of pulse amplitudes). The red dashed lines indicate the boundaries of the amplitude bins used for extracting pulse-to-pulse spectra. In each distribution, the total number of counts is evenly distributed among the bins.}
\label{distribution}
\end{figure}

\begin{figure}
\centering
\includegraphics[width=9cm]{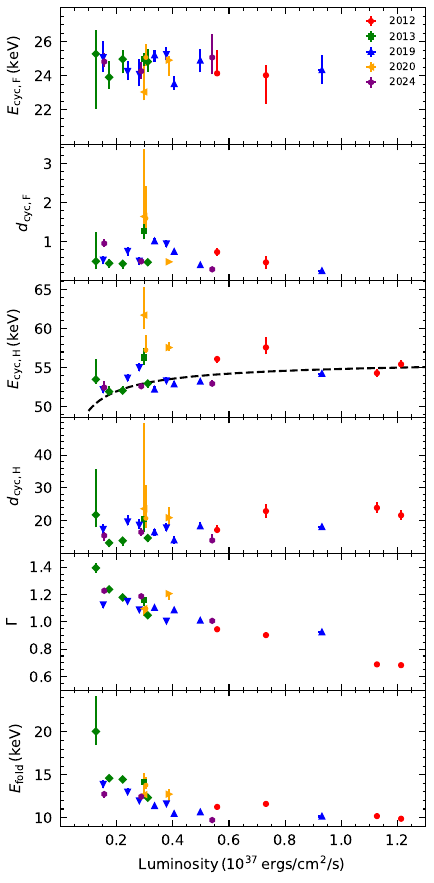}
\caption{Flux-resolved spectral parameters derived from the pulse-to-pulse analysis in the 3--70~keV energy range. {Different colors represent different observation years, and distinct marker styles indicate observation IDs; time-averaged results for the short-exposure observations (30002007002, 90602328002, 90602328004, and 90602328006) are also shown.} The dashed line in the second panel is the theoretical prediction for $E_* = 28.097 \pm 0.052$~keV (see equation \ref{eqt:E_theo}).}
\label{p2p_par_Lx}
\end{figure}
Following the method introduced by \citet{2011A&A...532A.126K}, we generated pulse amplitude-resolved spectra based on the pulse-to-pulse technique. This approach was applied to explore the luminosity dependence of the CRSF energy in Cep X–4 and V 0332+53 using \nustar data \citep{2017A&A...601A.126V,2018A&A...610A..88V}, and in 1A 0535+262 with \hxmt data \citep{2024MNRAS.528.7320S}. The pulse-to-pulse technique is based on the fact that most accreting pulsars exhibit strong pulse-to-pulse variations in amplitude that are driven by short-term fluctuations in the accretion rate. By selecting individual pulses with similar amplitudes and grouping them to generate spectra, this method extends the accessible luminosity range and improves statistics. Therefore, it enables a more detailed investigation of the luminosity dependence of the CRSF energy. 

The pulse amplitude is defined as the total number of counts summed over a single pulse. Due to limited photon statistics, extracting meaningful spectra from individual pulses is not feasible. Therefore, pulses with similar amplitudes were grouped together. We explored the distribution of pulse amplitudes and divided the entire range of amplitudes into several bins, thus ensuring approximately equal statistics within each bin. The number of bins was primarily determined by the available counting statistics for the energy spectra, which needed to be sufficient to constrain the cyclotron line energy during spectral fitting. For each bin a list of good time intervals (GTIs) was generated to select pulses within the corresponding amplitude range. Using these GTIs, we extracted the broadband spectra for each amplitude bin. 

Using the pulse-to-pulse method, we divided ObsIDs 10002007001, 30002007003, 90402339002, 30501003002, and 91002349002 into 4, 4, 4, 4, and 3 amplitude bins, respectively. The distributions of counts per individual pulse are shown in Figure \ref{distribution}. The distributions reveal that pulse amplitudes within a single observation cover a dynamic range of several times, indicating a strong intrinsic pulse-to-pulse variability of the source. For ObsIDs 30002007002, 90602328002, 90602328004, and 90602328006, due to limited exposure time, we retained only the time-averaged spectral results. For the spectral analysis, we employed the same model as used in the time-averaged analysis: \texttt{const$\times$tbabs$\times$pcfabs(gauss+gauss+gabs$\times$gabs$\times$fdcut)}. Given the shorter exposure time per pulse-to-pulse spectrum, we had to fix some parameters to their respective time-averaged values to achieve a good fit. We fixed the width of the iron line,  $\sigma_{\rm K\alpha}$; the energy and the width of the 10~keV line, $E_{\rm 10~keV}$ and $\sigma_{\rm 10~keV}$; and the width of the two CRSFs, $\sigma_{\rm cyc,F}$ and $\sigma_{\rm cyc,H}$, to their respective time-averaged values.  {We initially attempted to leave $\sigma_{\rm cyc,H}$ and $E_{\rm cut}$ free. However, this approach resulted in poor constraints on $E_{\rm cut}$. The continuum parameter $E_{\rm cut}$ was therefore fixed because of its degeneracy with the fundamental line energy $E_{\rm cyc,F}$. With the adopted strategy of fixing the above parameters, we obtained very good fits for all spectra.} 
The corresponding best-fit parameters are provided in the tables available at the CDS.

Figure \ref{p2p_par_Lx} presents the spectral results obtained from the pulse-to-pulse analysis. 
{The plot also includes the time-averaged results for the short-exposure observations 30002007002, 90602328002, 90602328004, and 90602328006.}
Our analysis extends the explored luminosity range to $(0.13\text{--}1.21) \times 10^{37}$\,erg\,s$^{-1}$ and successfully disentangles the intrinsic long-term evolution of $E_{\rm cyc}$ from its luminosity dependence.

The fundamental line energy, $E_{\rm cyc,F}$, exhibits no significant variation with time or luminosity, remaining stable at approximately 25~keV. At higher luminosities ($L_{\rm X} \gtrsim 10^{37}$\,erg\,s$^{-1}$), the spectra can be well fit without requiring an additional component for the fundamental line. Therefore, we do not include these two data points in the plot, as they are not meaningful when the corresponding spectral component is not necessary for the fit.
To improve the statistics, we split this observation into two bins, with only one data point in the high-luminosity bin. The fit remains acceptable without the fundamental line. When included, we derived a 3$\sigma$ upper limit on its strength of $\sim$ 0.8, suggesting that the line may be present but undetectable due to limited sensitivity.

For the energy of harmonic line, $E_{\rm cyc,H}$, excluding the 2020 observations, {the data appear fairly stable, with slightly lower values at lower luminosities.} Moreover, at comparable luminosity levels, there is no time dependence of the $E_{\rm cyc,H}$ across different epochs, suggesting that the long-term decay of $E_{\rm cyc,H}$ likely ceased after 2012.
Additionally, the photon index exhibits a negative correlation with luminosity, indicating a hardening of the continuum spectrum at higher flux levels.

\section{Discussion} \label{sec:discussion}

{In this work, we investigate the dependence of the fundamental cyclotron line energy ($E_{\mathrm{cyc,F}}$) and the harmonic line energy ($E_{\mathrm{cyc,H}}$) on both time and X-ray luminosity ($L_{\rm X}$). By performing a time-averaged analysis of archival \bat monitoring data, we investigated the long-term evolution of $E_{\mathrm{cyc,H}}$ over 20 years. Using observations from \nustar we performed the pulse-to-pulse analysis to reveal the comprehensive scenario of the $E_{\rm cyc}$--$L_{\rm X}$ relation covering a luminosity range of 0.13 $\times$ $10^{37}\,$erg\,$\rm{s}^{-1}$ to 1.21 $\times$ $10^{37}\,$erg\,$\rm{s}^{-1}$.}

\subsection{Changes in $E_{\rm cyc}$ with time}

To date, only two sources, Her X-1 and Vela X-1, have shown clear long-term variability in $E_{\rm cyc}$ over timescales of tens of years \citep{2020A&A...642A.196S, 2016MNRAS.463..185L, 2019MNRAS.484.3797J}. 
In Her X-1 the long-term decay of $E_{\rm cyc}$ ended around 2012, stabilizing at approximately 37 keV, with a decline rate of $\sim$0.26 keV yr$^{-1}$ between 1996 and 2012 \citep{2020A&A...642A.196S}.
For Vela X-1 our analysis confirms the previously reported decrease in the energy of the first cyclotron harmonic \citep{2016MNRAS.463..185L, 2019MNRAS.484.3797J}, which now appears to remain constant. The decrease rate is $\sim$ 0.36\,keV\,$\rm{yr}^{-1}$ and $\sim$ 0.51\,keV\,$\rm{yr}^{-1}$ stated by \citet{2016MNRAS.463..185L} and \citet{2019MNRAS.484.3797J}, similar to our result $\sim$ 0.58\,keV\,$\rm{yr}^{-1}$ based on BAT observations. 

The long-term decay of $E_{\rm cyc,H}$ in \src is likely a local effect at the magnetic polar cap, rather than a change in the global dipole field. 
Similar scenarios have been proposed for Her X-1 \citep{2014A&A...572A.119S, 2016A&A...590A..91S, 2020A&A...642A.196S}. They suggested that the decay could be connected to a geometric displacement of the emission region in the dipole field or it could be related to the accreted matter that accumulates into a magnetically supported mound and causes a change in the magnetic field configuration at the polar cap. The end of the decay could be explained by the mound reaching its maximum stable size beyond which further accumulation no longer significantly affects the field structure.

In addition, we observed a sudden increase in the harmonic cyclotron line energy, reaching $\sim$62~keV in 2020, with \nustar observations, followed by a decrease back to $\sim$52~keV in 2024 {at comparable luminosity levels of $\sim$$0.3 \times 10^{36}\,$erg\,$\rm{s}^{-1}$}. Such a dramatic evolution in the cyclotron line energy is unusual. This result is supported by \bat monitoring, which shows a significant increase in $E_{\rm cyc,H}$ between 2020 and 2023 (MJD 59000–60000). A similar hump around MJD 55000 was also detected in the \bat data and was reported by \citet{2019MNRAS.484.3797J}. A comparable phenomenon was observed in Her X-1, where the line energy remained constant prior to 1991 and then increased to $\sim$41~keV between 1990 and 1995 \citep{1999hxra.conf...33G, 1998A&A...329L..41D}, followed by a more gradual, linear decay. This process may be cyclic \citep{2017A&A...606L..13S}.

We present here a possible explanation for the sudden increase. Matter accumulation at the magnetic poles increases the overpressure until a critical threshold is reached. This triggers interchange instabilities (e.g., ballooning instabilities \citep{2001ApJ...553..788L}) near the column boundary, causing localized magnetic field deformation and mass leakage. As a result, $E_{\rm cyc,H}$ increases because of partial column collapse, and a reduction in height. The fact that only $E_{\rm cyc,H}$ exhibits this strong variability, while the $E_{\rm cyc,F}$ remains nearly constant, may suggest that the harmonic line is more sensitive to changes in the height of the emission region.

It is worth noting that previous studies have shown that sudden changes in cyclotron line energy generally occur during the declining phase of long‑term evolution. In contrast, this work is the first to reveal that such abrupt shifts can also arise when the cyclotron line has entered a relatively stable, long‑term plateau. This finding suggests that even when the polar‑cap accretion structure appears to have reached a steady state, localized regions may still experience dynamic or structural changes. At present, no comprehensive theoretical framework can fully explain this behavior, but extensive future observations of Vela X‑1 and other source (e.g. Her X-1, Cen X-3, 4U 1538-522) could provide critical constraints for refining models of polar‑cap dynamics and magnetic‑field evolution.

\subsection{Changes in $E_{\rm cyc}$ with luminosity}

The variation in $E_{\rm cyc}$ with luminosity has long been a key focus in the study of accreting X-ray pulsars because it directly reflects the strength of the local magnetic field. Previous observations suggest that cyclotron lines exhibit two distinct evolutionary trends depending on luminosity. A positive dependence of $E_{\rm cyc}$ on $L_{\rm X}$ was observed in low- to moderate-luminosity sources, such as Her X-1 \citep{2007A&A...465L..25S,2014A&A...572A.119S, 2016A&A...590A..91S, 2017A&A...606L..13S, 2020A&A...642A.196S}, \src \citep{2014AAS...22343820F, 2016MNRAS.463..185L, 2022A&A...660A..19D}, and Cep X-4 \citep{2017A&A...601A.126V}. 
A negative dependence of $E_{\rm cyc}$ on $L_{\rm X}$ was observed in relatively high-luminosity sources, such as V 0332+53 \citep{2006MNRAS.371...19T,2016MNRAS.460L..99C, 2017MNRAS.466.2143D, 2018A&A...610A..88V} and 1A 0535+262 \citep{2021ApJ...917L..38K}. 
The luminosity at which these different evolutionary behaviors of $E_{\rm cyc}$ occur is often referred to as the critical luminosity ($L_{\rm crit}$) of the source, typically a few times $10^{37}\,$erg\,$\rm{s}^{-1}$. The $L_{\rm crit}$ separates the accretion regimes into sub-critical and super-critical states. In general, sub-critical accretion is associated with a positive $E_{\rm cyc}$–$L_{\rm X}$ correlation, while super-critical accretion corresponds to a negative correlation \citep{2012A&A...544A.123B, 2015MNRAS.447.1847M}. The contrasting $E_{\rm cyc}$–$L_{\rm X}$ behaviors at sub-critical and super-critical luminosities can be interpreted within different theoretical models. An interpretation for the observed $E_{\rm cyc}$–$L_{\rm X}$ correlation relies on the relationship between the $E_{\rm cyc}$ and the height of the emission region above the NS surface \citep{2012A&A...544A.123B}.  In low- to moderate-luminosity sources, including \src, the accreting material is primarily decelerated via Coulomb braking. As the accretion rate increases, the emission region is pushed closer to the NS surface, where the magnetic field is stronger, leading to an increase in $E_{\rm cyc}$. \citet{2015MNRAS.454.2714M} proposed that the positive correlation observed in sub-critical sources arises from Doppler shifts in the infalling plasma. In this model, CRSFs are formed as radiation from the polar-cap hotspot travels upward through the accretion column and interacts with the plasma at resonant energies. As luminosity increases, radiation pressure slows the plasma, thus reducing the Doppler redshift and causing the observed line energy to rise.

In our study we first focused on the harmonic cyclotron line energy, $E_{\rm cyc,H}$. Using two \nustar observations from 2012 and 2013, \citet{2014AAS...22343820F} reported a positive correlation between $E_{\rm cyc,H}$ and luminosity. However, \citet{2022A&A...660A..19D}, based on two \nustar observations from 2019, were unable to confirm this correlation. 
{According to our results, the data remain relatively stable, exhibiting modestly reduced values at lower luminosities.}
Theoretical predictions for the $E_{\rm cyc}$–$L_{\rm X}$ relationship can be derived from Equations (51) and (58) of \citet{2012A&A...544A.123B}:

\begin{equation}
\begin{aligned}
E_{\rm theo} = & \left[1 + 0.6 \left(\frac{R_*}{10\,\text{km}}\right)^{-\frac{13}{14}}\left(\frac{\Lambda}{0.1}\right)^{-1}\left(\frac{\tau_*}{20}\right) \times\left(\frac{M_*}{1.4\,M_{\odot}}\right)^{\frac{19}{14}}\right.\\
&\left.\left(\frac{E_*}{1\,\text{keV}}\right)^{-\frac{4}{7}}\left(\frac{L_x}{10^{37}\,\text{erg\,s}^{-1}}\right)^{-\frac{5}{7}}\right]^{-3}\times E_*.
\end{aligned}
\label{eqt:E_theo}
\end{equation}
Here $\tau_*$ is the Thomson optical depth, around 20 for typical HMXB parameters \citep{2012A&A...544A.123B}, and $E_*$ is the energy of the fundamental cyclotron line at the NS surface. We adopt the values $\Lambda$=1, $R_*$=10~km and $M_*$=1.8 $M_{\odot}$. Fitting the data using Equation~\ref{eqt:E_theo} and excluding the anomalously high values from 2020 yields $E_* = 28.097 \pm 0.052$~keV, assuming a harmonic ratio of 2. The Pearson correlation coefficient is 0.462, with a p-value of 0.0405.

However, the fundamental line shows a different behavior as a function of luminosity. A possible anti-correlation up to $7 \times 10^{36}\,$erg\,$\rm{s}^{-1}$ was reported by \citet{2014AAS...22343820F}, beyond which the correlation seems to flatten. This result is obtained from the fitting of data from each orbit, with relatively short exposure times, which leads to large error bars. In our pulse-to-pulse analysis, we considered both the spectra with similar pulse amplitudes and those with longer exposures. As a result, our findings provide better constraints on the fundamental cyclotron line and confirm that its energy shows a flat evolutionary trend with luminosity. 
The lack of a clear luminosity dependence in the fundamental CRSF energy can be explained by the photon spawning effect \citep{2007A&A...472..353S}. In this scenario, resonant scattering events produce additional low-energy photons that preferentially populate the energy range of the fundamental line, effectively filling in the absorption feature and diminishing its depth. As a result, the fundamental line becomes less sensitive to luminosity. 
Furthermore, based on \citet{1991ApJ...374..687H}, the cross section of CRSF at the fundamental and the harmonic is different, which is related to the angle between the photon momentum and the magnetic field. In the subcritical regime, photons can escape from the top of the accretion column along the magnetic field lines (called pencil-beam geometry), where the cross section of the harmonic line is much lower than that of the fundamental, making it intrinsically weaker. However, when the infalling material reaches relativistic speeds, the resulting beaming effect can increase the effective viewing angle and enhance the cross section of the harmonic line. Such conditions may arise near the shock region of the accretion column, leading to a locally stronger harmonic feature (see discussion in \citet{2022ApJ...932..106K}).

To further investigate the accretion regime of \src, we calculated the $L_{\mathrm{crit}}$ under different theoretical models.
The expression for the $L_{\mathrm{crit}}$ as a function of the magnetic field strength at the neutron star (NS) surface, following \citet{2012A&A...544A.123B}, is given by:
\begin{equation}
\begin{aligned}
L_{\mathrm{crit}} =\; & 1.49 \times 10^{37}~\mathrm{erg\,s^{-1}} 
\left( \frac{\Lambda}{0.1} \right)^{-7/5} 
w^{-28/15} \\
& \times \left( \frac{M_*}{1.4 M_\odot} \right)^{29/30}
\left( \frac{R_*}{10\,\mathrm{km}} \right)^{1/10}
\left( \frac{B_*}{10^{12}\,\mathrm{G}} \right)^{16/15},
\end{aligned}
\label{eqt:L_crit}
\end{equation}
where $R_*$ is the radius of the NS, $M_*$ is the mass of the NS, $B_*$ is the magnetic field strength at the NS surface, $\Lambda$ is a constant that depends on the accretion flow geometry, and $w$ is a parameter describing the spectral shape inside the column. We adopt $R_* = 10$~km and $M_* = 1.8~M_{\odot}$ \citep{2011ApJ...730...25R}, and assume $w = 1$, corresponding to a Bremsstrahlung-dominated emission spectrum inside the column. Accretion flow geometry is a source of principal uncertainties. If we take $\Lambda = 1$, appropriate for spherical or wind-fed accretion, and combine Equation~\ref{eqt:L_crit} with Equation~\ref{equ:Bcyc}, assuming a surface fundamental cyclotron line energy of $E_{\rm cyc,F} = 25$~keV, we obtain $L_{\mathrm{crit}} \sim 0.17 \times 10^{37}$~erg~s$^{-1}$. If we adopt $\Lambda = 0.1$, which corresponds to disc accretion, yields $L_{\mathrm{crit}} \sim 4.31 \times 10^{37}$~erg~s$^{-1}$. These two scenarios represent limiting cases. In reality, the accretion flow may involve a combination of both modes, suggesting that the actual $L_{\mathrm{crit}}$ lies between these two limits. 

\citet{2015MNRAS.447.1847M} provided a numerical solution by calculating the luminosity for two scenarios: one with purely extraordinary polarization and another with an equal mix of ordinary and extraordinary polarization. The real critical luminosity value is expected to fall between the two cases. In their prediction, assuming $E_{\rm cyc}$ = 25~keV, the estimated $L_{\mathrm{crit}}$ for \src is in the range of $\sim 0.3-1.0 \times 10^{37}$~erg~s$^{-1}$ (as shown in  Fig. 7 of that study).

We observe a clear spectral hardening with increasing luminosity. In many accreting pulsars, the photon index $\Gamma$ exhibits a luminosity dependence similar to that of the $E_{\rm cyc}$: it decreases with increasing luminosity in high-luminosity sources (i.e., the spectrum softens), and increases with luminosity in low- to moderate-luminosity sources (i.e., the spectrum hardens) \citep{2011A&A...532A.126K, 2013A&A...551A...1R}. In the case of \src, this behavior can be interpreted as a result of the emission region moving closer to the neutron star surface at higher luminosities. As the sinking region becomes more compact, the optical depth increases, leading to enhanced Comptonization and the production of harder photons, thereby yielding a harder spectrum \citep{2012A&A...544A.123B, 2013A&A...551A...1R}. According to the model proposed by \citet{2015MNRAS.452.1601P}, at low luminosities the emerging X-ray spectrum is produced by ordinary photons in a magnetized optically thin slab-like atmosphere near the polar cap. As the accretion rate increases, the Comptonization rises, thus leading to a harder X-ray continuum.

\section{Conclusions} \label{sec:conclusions}

{We conducted a detailed study of the long-term evolution and the luminosity dependence of the cyclotron line energy in \src, based on archival \bat monitoring and nine \nustar observations from 2012 to 2024. }

This work presented the first confirmation that the long-term decay of the harmonic line energy in \src has ended. Moreover, we report the first detection of a renewed increase in $E_{\rm cyc,H}$ between 2020 and 2023. Such dramatic variability, which occurred after the line energy had settled into a relatively stable plateau, is unusual and likely reflects sudden magnetic or structural changes in the accretion environment. Continued long-term monitoring of \src and other sources will be essential to determine whether this behavior is part of a broader cyclic phenomenon. 

Using pulse-to-pulse analysis, we find that the fundamental line energy remains stable over both time and luminosity, showing no significant evolution. {The harmonic line energy is found to be slightly lower at low luminosities and to flatten at higher luminosities.}

\section*{Data availability}
The tables related to the \bat spectral parameters, the \nustar averaged spectral parameters,
and the \nustar pulse-to-pulse spectral parameters for individual observations
are available in electronic form at the CDS via anonymous ftp to
cdsarc.u-strasbg.fr (130.79.128.5) or via
\url{http://cdsweb.u-strasbg.fr/cgi-bin/qcat?J/A+A/}.

\begin{acknowledgements}
This work made use of data from the \nustar mission, a project led by the California Institute of Technology, and also benefited from the long-term support of the Burst Alert Telescope on board the Neil Gehrels {\it Swift} observatory. Y.J. Du would like to thank the support from China Scholarship Council (CSC 202108080247). P. J. Wang is grateful for the financial support provided by the Sino-German (CSC-DAAD) Postdoc Scholarship Program (57678375). L. Ji is supported by the National Natural Science Foundation of China under grant No. 12173103. LD acknowledges funding from the Deutsche Forschungsgemeinschaft (DFG, German Research Foundation) - Projektnummer 549824807.
\end{acknowledgements}

\bibliographystyle{yahapj}
\bibliography{references}

\clearpage
\onecolumn

\begin{appendix}

\section{Additional figure}\label{sec:appendix}

\begin{center}
\includegraphics[width=19cm]{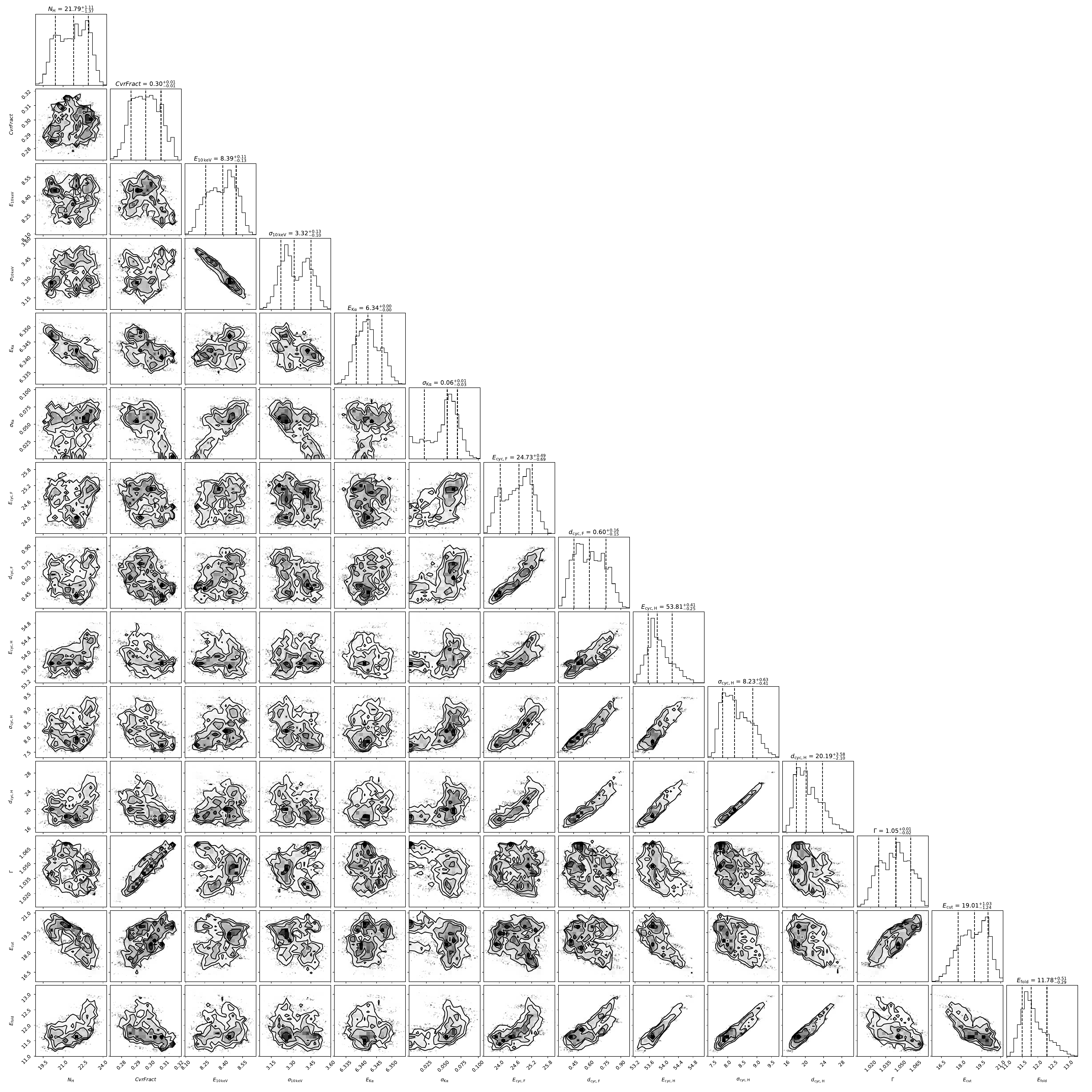}
\end{center}
\vspace{-1em}
\captionof{figure}{Corner plot of the posterior distributions of all model parameters for ObsID 91002349002.}
\label{Corner_plot}

\end{appendix}
\end{document}